\documentclass{jpp}
\usepackage{graphicx}
\usepackage{epstopdf, epsfig}

\usepackage{subcaption}
\usepackage{amsmath}
\usepackage{amssymb}
\usepackage{color}
\usepackage{xspace}

\newcommand{\ov}[1]{\overline{#1}}
\newcommand{\be}{
\begin{equation}
}

\newcommand{\ee}{
\end{equation}
}

\newcommand{\ba}{\begin{eqnarray}}
\newcommand{\ea}{\end{eqnarray}}
\newcommand{\bas}{\begin{eqnarray*}}
\newcommand{\eas}{\end{eqnarray*}}

\newcommand{\vc}[1]{
{\bf{#1}}
}

\newcommand{\intl}{
\int \limits
}

\newcommand{\df}{
{\rm d}
}

\newcommand{\wt}[1]{
\widetilde{#1}
}

\newcommand{\totd}[2]{
\frac{\df #1}{\df #2}
}

\newcommand{\eps}{\epsilon}

\newcommand{\gav}[1]{
\langle #1 \rangle
}
\newcommand{\Bgav}[1]{
\Big\langle #1 \Big\rangle
}
\newcommand{\kp}{\kappa}
\newcommand{\G}{\Gamma}

\newcommand{\ohatTwo}{\tilde{\omega}_d}
\newcommand{\osT}{\omega_T}
\newcommand{\osn}{\omega_n}

\newcommand\ie{\textit{i.e.}\xspace}

\renewcommand\Re{\mbox{$\mathrm{Re}$}}
\renewcommand\Im{\mbox{$\mathrm{Im}$}}

\def \kperp {k_{\perp}}

\newcommand{\xpar}{\ensuremath{x_{\parallel}}}
\newcommand{\xperp}{\ensuremath{x_{\perp}}}

\newcommand{\avg}[1]{\overline{#1}}

\shorttitle{Electrostatic stability of electron-positron plasmas}
\shortauthor{A.~Mishchenko, G.~Plunk, and P.~Helander}

\title{Electrostatic stability of electron-positron plasmas in dipole geometry}

\author{Alexey Mishchenko\aff{1}
  \corresp{\email{alexey.mishchenko@ipp.mpg.de}}, 
  Gabriel G. Plunk\aff{1}
  \and Per Helander\aff{1} 
  }

\affiliation{\aff{1}Max Planck Institute for Plasma Physics,
  D-17491 Greifswald, Germany}

\begin{document}

\maketitle

\begin{abstract}
The electrostatic stability of electron-positron plasmas is investigated in the point-dipole and Z-pinch limits of dipole geometry.  
The kinetic dispersion relation for sub-bounce-frequency instabilities is derived and solved.  For the zero-Debye-length case, the stability 
diagram is found to exhibit singular behavior.  However, when the Debye length is non-zero, a fluid mode appears,
which resolves the observed singularity, and also demonstrates that both the temperature and density gradients can drive instability.
It is concluded that a finite Debye length is necessary to determine the stability boundaries in parameter space.
Landau damping is investigated at scales sufficiently smaller than the Debye length, where instability is absent.
\end{abstract}

\section{Introduction}

The prospects of creating magnetically confined electron-positron (pair) plasmas 
in dipole or stellarator geometries have been discussed since early 2000's
\citep{pedersen2003}. In the near future, the first experiment will
be constructed to confine such plasmas in magnetic dipole geometry \citep{pedersen2012}. 
Recently, efficient injection and
trapping of a cold positron beam in a dipole magnetic field configuration 
has been demonstrated by \citet{saitoh2015} using a supported permanent magnet. 
This result is a key step towards the further studies using the levitated magnetic coil with the 
ultimate aim of creating and studying the first man-made
magnetically-confined pair plasma in the laboratory.

It has been shown by \citet{per_prl} that pair plasmas possess unique
stability properties due to the mass symmetry between the
particle species. For example, drift instabilities are completely absent
in straight-field-line geometry, e.~g.~in a slab, provided that the temperature and density profiles 
of the two species are equal (``symmetric'' pair plasmas). The symmetry between the species 
is broken if the temperature profiles of the electrons and positrons differ 
or there is an ion contamination. 
Then, the drift instabilities can be excited \citep{Mishchenko_slab} even in unsheared slab geometry. 
In a sheared slab, pure pair plasmas are prone to current-driven reconnecting instabilities \citep{Zocco_alf},
although asymmetry between the species is also needed in this case since the ambient electron
flow velocity must differ from the positron one for the ambient current to be finite.
In contrast to slab geometry, a dipole magnetic field has finite curvature. In this case, 
the symmetry between the species is broken by the curvature drifts and the plasma is driven unstable  
by the temperature and density gradients \citep{per_prl}, 
even without ion contamination and for identical temperature profiles of the two species. 
This result persists also in the electromagnetic regime \citep{per_jpp}. The nonlinear stability of dipole pair plasmas 
has also been addressed recently by \citep{per_spineto}. 

In this paper, we extend the results of \citet{per_prl} by performing a detailed 
study of the drift-kinetic stability of pure pair plasma in dipole
geometry, making use of both the Z-pinch and point-dipole limits, where the dispersion relation is derived and numerically solved.
The structure of the paper is as follows. In \S\ref{Dipole_field}, we introduce the magnetic dipole field 
and discuss the near-magnetic-axis (Z-pinch) and far-field (point-dipole) limits. 
In \S\ref{GK_Theory}, we introduce the linear drift-kinetic description and derive 
a ``master'' equation, applicable to both Z-pinch and point-dipole limits.  The Z-pinch and point-dipole
limits are then individually treated in \S\ref{sec:Z_pinch} and \S\ref{sec:dipole}, respectively.
The conclusions are summarised in \S\ref{sec:conclusions}. 
%
%
\section{Dipole magnetic field} \label{Dipole_field}
In cylindrical coordinates $(r,\varphi,z)$, the magnetic field of a
circular conducting loop with the radius $r_0$ carrying the total current $I$ is
\be
\vc{B}(r,z) = \nabla \psi(r,z) \times \nabla \varphi \ , \;\;\; 
\nabla \varphi = \frac{\vc{e}_{\varphi}}{r} 
\ee
with the poloidal magnetic flux given by \citep{Landau,Simpson:dipole}
\be
\psi(r,z) = \frac{C}{2} \, \sqrt{(r_0 + r)^2 + z^2} \, \left[
  \frac{r_0^2 + r^2 + z^2}{(r_0 + r)^2 + z^2} K(\kp) - E(\kp) \right] \ , \;\;
C = \frac{\mu_0 I}{\pi}\label{psi-eqn}
\ee
defined in terms of the elliptic integrals of the first and the second kind:
\be
K(k) = \intl_0^1 \frac{\df x}{\sqrt{(1 - x^2) (1 - k^2 x^2)}} \ , \;\;
E(k) = \intl_0^1 \frac{\sqrt{1 - k^2 x^2}}{\sqrt{1 - x^2}} \; \df x
\ee
Following \citet{Simpson:dipole}, we define
\be
\kappa^2 = 1 - \rho^2/\beta^2 \ , \;\;\;
\rho^2 = (r - r_0)^2 + z^2 \ , \;\;\;
\beta^2 = (r_0 + r)^2 + z^2.
\ee
The components of the magnetic field are then expressed
\ba
\label{B_r}
B_r &=& \frac{C z}{2 \rho^2 \beta r} \left[(r_0^2 + r^2 + z^2)E(\kappa^2) -
  \rho^2 K(\kappa^2) \right] \\
\label{B_z}
B_z &=& \frac{C}{2 \rho^2 \beta} \left[(r_0^2 - r^2 - z^2)E(\kappa^2) + 
  \rho^2 K(\kappa^2) \right] 
\ea 

\begin{figure}
\centering
\includegraphics[width=0.4 \textwidth]{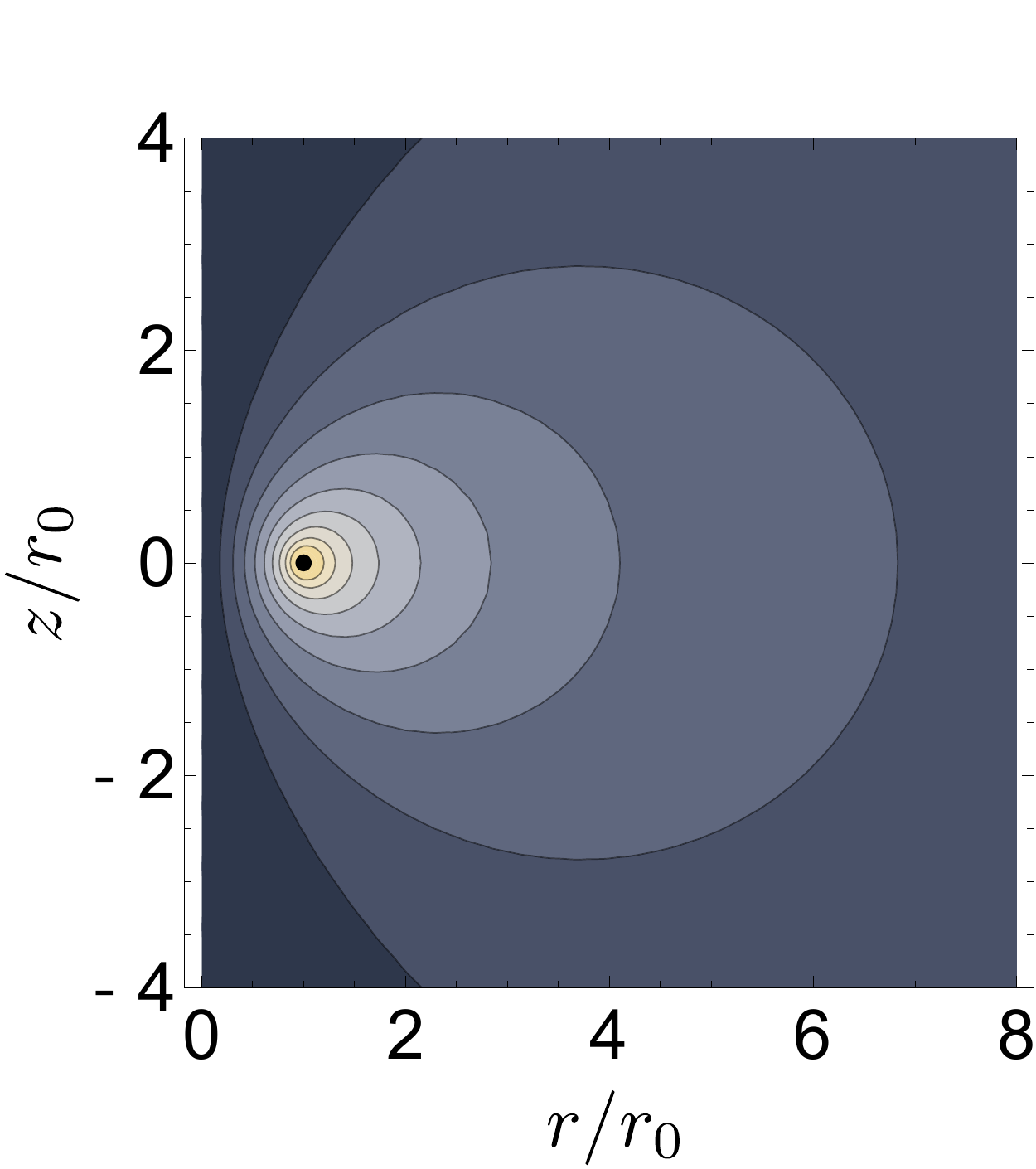}
\includegraphics[width=0.4 \textwidth]{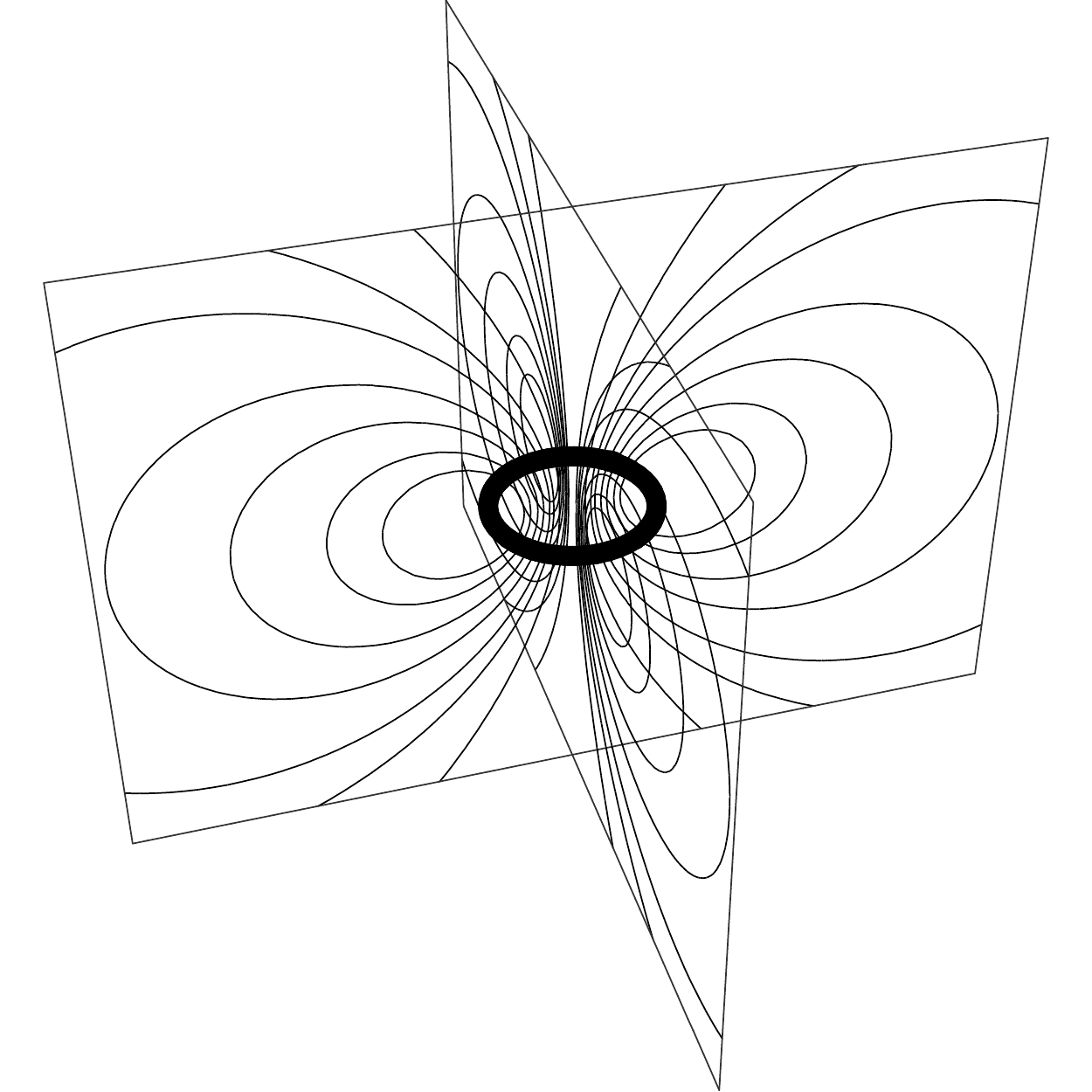}
\caption{On the left is a plot of $\psi(r, z)$, as defined in Eqn.~\ref{psi-eqn}.  Magnetic field lines (here, represented by contours of constant $\psi$) are generated by the circular coil at $(r, z) = (r_0, 0)$, whose cross section is shown here as a small black disk. In the yellow region, close to the coil, the field is approximately that of a Z-pinch, whereas in the grey region, far from the coil, it approaches the field of a point dipole.  On the right is a 3D cartoon of a dipole magnetic field.} 
\label{dipole-geometry-fig}%
\end{figure}

We will consider two asymptotic cases of this magnetic geometry, depicted in Fig.~\ref{dipole-geometry-fig}.  The first case arises in the region close to the current loop.  This is the Z-pinch limit, corresponding to 
\be
\frac{r - r_0}{r_0} \sim \frac{z}{r_0} \sim \frac{\rho}{r_0} \ll 1.
\ee
In this case, $\beta \approx 2 r_0$ 
so that $\kappa \rightarrow 1$ \citep{gradshteyn} and 
\be
E(\kappa) \approx 1 \ , \;\; 
K(\kappa) \approx \ln\left(\frac{4}{\sqrt{1 - \kappa^2}}\right) 
\approx \,-\,\ln\left(\frac{\rho}{8 r_0}\right) 
\ee
Substituting these relations into Eqs.~(\ref{B_r}) and (\ref{B_z}), we obtain
\be
B_r \approx \frac{C z}{2 \rho^2}   \ , \;\; 
B_z \approx \,-\,\frac{C (r - r_0)}{2 \rho^2} \ , \;\;
\psi = \,-\, \frac{C r_0}{2} \ln\left(\frac{\rho}{8 r_0}\right)  
\ee
Here, we employed $\rho \ll r_0$ and invoked
\be
\lim_{x \rightarrow 0} x \ln x = 0 \; \Rightarrow \;
\frac{\rho}{r_0} \, \ln\left(\frac{\rho}{r_0}\right) \ll 1 
\ee
Introducing a new ``quasi-polar'' coordinate system, with the axis
corresponding to the magnetic axis (the current loop), we define the new
``quasi-polar'' angle $\zeta$ satisfying
\be
\sin\zeta = \frac{z}{\rho} \ ,  \;\;
\cos\zeta = \frac{r - r_0}{\rho} 
\ , \;\; \vc{e}_{\zeta} = \frac{\nabla \zeta}{|\nabla \zeta|} = 
\,-\, \vc{e}_r \sin \zeta + \vc{e}_z \cos\zeta
\ee
In these notations, the magnetic field becomes
\be
B_r = \frac{\mu_0 I}{2 \pi \rho} \sin\zeta \ ,\;\;
B_z = \,-\,\frac{\mu_0 I}{2 \pi \rho} \cos\zeta 
\ , \;\;
\vc{B} = \,-\,\frac{\mu_0 I}{2 \pi \rho} \, \vc{e}_{\zeta}
\ee
This is the usual Z-pinch magnetic field, created by a linear current flowing in the axial direction.

The second case to be considered corresponds to the far-field limit. For this
case, the spherical coordinate system $(\hat{r}, \hat{\theta}, \varphi)$, with $\hat{r}$ the
spherical radial distance, $\hat{\theta}$ the azimuthal angle, and $\varphi$ the polar angle, is
more convenient. For $\hat{r} \gg r_0$, the magnetic field is  
\be
B_r \approx \frac{2 M \cos \hat{\theta}}{\hat{r}^3} \ , \;\; 
B_{\theta} \approx \frac{M \sin \hat{\theta}}{\hat{r}^3} 
\ , \;\; \psi = \frac{M \sin^2 \hat{\theta}}{\hat{r}}
\ ,\;\; M = \frac{\mu_0 I r_0^2}{4}
\ee
These expressions are called the ``point-dipole approximation'', valid far from 
the current loop.
%
%
\section{Drift-kinetic theory} \label{GK_Theory}
Following \citet{per_prl} and \citet{per_jpp}, we begin with gyrokinetic theory. It is convenient to write
the gyrokinetic distribution function in the form:
\be
f_a = f_{a0} \left(1 - \frac{e_a \phi}{T_a}\right) + g_a = f_{a0} + f_{a1} \ , \;\;
f_{a1} = \,-\,\frac{e_a \phi}{T_a}\,f_{a0} + g_a 
\ee
Here, $f_{a0}$ is a Maxwellian, $a$ is the species index with $a=e$
corresponding to the electrons, $a = p$ to the positrons. The linearised gyrokinetic equation in this notation is 
\be
i v_{\|} \nabla_{\|} g_a + (\omega - \omega_{{\rm d}a}) g_a  = \frac{e_a}{T_a}
\, J_0\left(\frac{k_{\perp}v_{\perp}}{\omega_{{\rm c}a}}\right) \, 
\Big(\omega - \omega^T_{*a}\Big) \, \phi \, f_{a0}\label{gk-eqn}
\ee
with $J_0$ the Bessel function, $\omega_{{\rm c}a}$ the cyclotron frequency,
$k_{\perp}$ the perpendicular wave number, and $\phi$ the perturbed electrostatic
potential. The notation used is summarized as follows:
\ba
&&{} \omega_{*a}^T = \omega_{*a} \left[ 1 + 
\eta_a \left( \frac{v^2}{v_{{\rm th}a}^2} - \frac{3}{2}\right) \right] \ , \;\;
v = \sqrt{v_{\|}^2 + v_{\perp}^2} \ , \;\; \mu = \frac{m_a v_{\perp}^2}{2 B} \\
&&{} \omega_{*a} = \frac{k_{\varphi} T_a}{e_a} \totd{\ln n_a}{\psi} \ , \;\;
\eta_a = \totd{\ln T_a}{\ln n_a} \ , \;\; 
v_{{\rm th}a} = \sqrt{\frac{2 T_a}{m_a}}  \ , \;\; \omega_{{\rm c}a} = \frac{e_a B}{m_a} \\
&&{} \omega_{{\rm d}a} = \vc{k}_{\perp} \cdot \vc{v}_{{\rm d}a}   \ , \;\; 
\vc{v}_{\rm d} = \Big(m v_{\|}^2 + \mu B \Big)  \, \frac{\vc{b} \times \nabla B}{q B^2}
\ , \;\; \vc{k}_{\perp} = k_{\psi} \nabla \psi + k_{\varphi} \nabla \varphi
\ea
Here, $\psi$ is the poloidal flux and $\varphi$ is the polar (toroidal) angle. 
We choose the sign convention such that $\omega_{*e} \ge 0$ for the electrons and $\omega_{*p} \le 0$ 
for the positrons. We will assume the drift-kinetic limit in what follows, 
\ie $k_\perp v_{{\rm th}a}/\omega_{{\rm c}a} \ll 1$ so $J_0 \approx 1$.

Applying the bounce average to Eqn.~\ref{gk-eqn}, we obtain to lowest order 
\be
(\omega - \avg{\omega}_{{\rm d}a}) g_a = (\omega - \omega_{\ast a}^T) \frac{e_a \avg{\phi}}{T} f_{a0},
\ee
with the bounce-average operation defined as
\be
\ov{(\ldots)} = \oint (\ldots) \frac{\df l}{v_{\|}} \Big/ \oint \frac{\df l}{v_{\|}}.
\ee
Here, $l$ is the arc length measured along a magnetic field line and the integration is performed between 
bounce points for trapped particles, and over the entire closed field line for passing particles. 
Note that there are only trapped particles in the point-dipole limit and only passing particles in the Z pinch limit.
We assume the temperature and the density profiles of the electrons and the positrons to be identical, 
and invoke the Poisson equation:
\be
\left( \sum_{a=e,p} \frac{n_a e_a^2}{T_a} + \eps_0 \, k_{\perp}^2 \right) \phi
= \sum_{a=e,p} e_a \int g_a \df^3 v \ , \;\;
\ee
We find that the perturbed electrostatic potential satisfies the equation:
\be
\label{gk_per}
\Big( 1 + k_{\perp}^2 \lambda_D^2 \Big) \phi = \frac{1}{n_0} \int
\frac{\omega^2 - \ov{\omega}_d \omega_*^T}{\omega^2 - \ov{\omega}_d^2}
\ov{\phi} \, f_0 \df^3 v 
\ee
Here and in the following, we use the notation $\omega_*^T \equiv \omega_{*e}^T$, 
$\omega_* \equiv \omega_{*e}$, $\ov{\omega}_d \equiv \ov{\omega}_{{\rm d}e}$, $n_0 = n_e$, 
$T_0 \equiv T_e$, and the Debye length is defined as usual $\lambda_D = \sqrt{\eps_0 T_0 / (2 n_0 e^2)}$.

Eq.~(\ref{gk_per}) is the ``master'' equation for drift-kinetic stability in magnetic dipole geometry. 
It will be solved in Z-pinch and point-dipole limits. This will give us insight into the general properties of the 
stability of symmetric pair plasmas in magnetic dipole geometry.
%
%
\section{Z-pinch case} \label{sec:Z_pinch}
In the Z-pinch limit, the components (in polar coordinates) of the magnetic field and the perpendicular wave vector, i.~e.~$\vc{B}\cdot \vc{e}_{\zeta}$ and $\vc{k}_\perp\cdot \vc{e}_{\zeta}$, etc.~ are flux functions. In this case, there is no particle trapping, and the orbit average of the perturbed electrostatic potential $\bar{\phi}$ coincides with its field-line average $\gav{\phi}$
\be
\gav{\phi} = \oint \phi \, \df l \Big/ \oint \df l
\ee

\begin{figure}
\centering
\includegraphics[width=0.75 \textwidth]{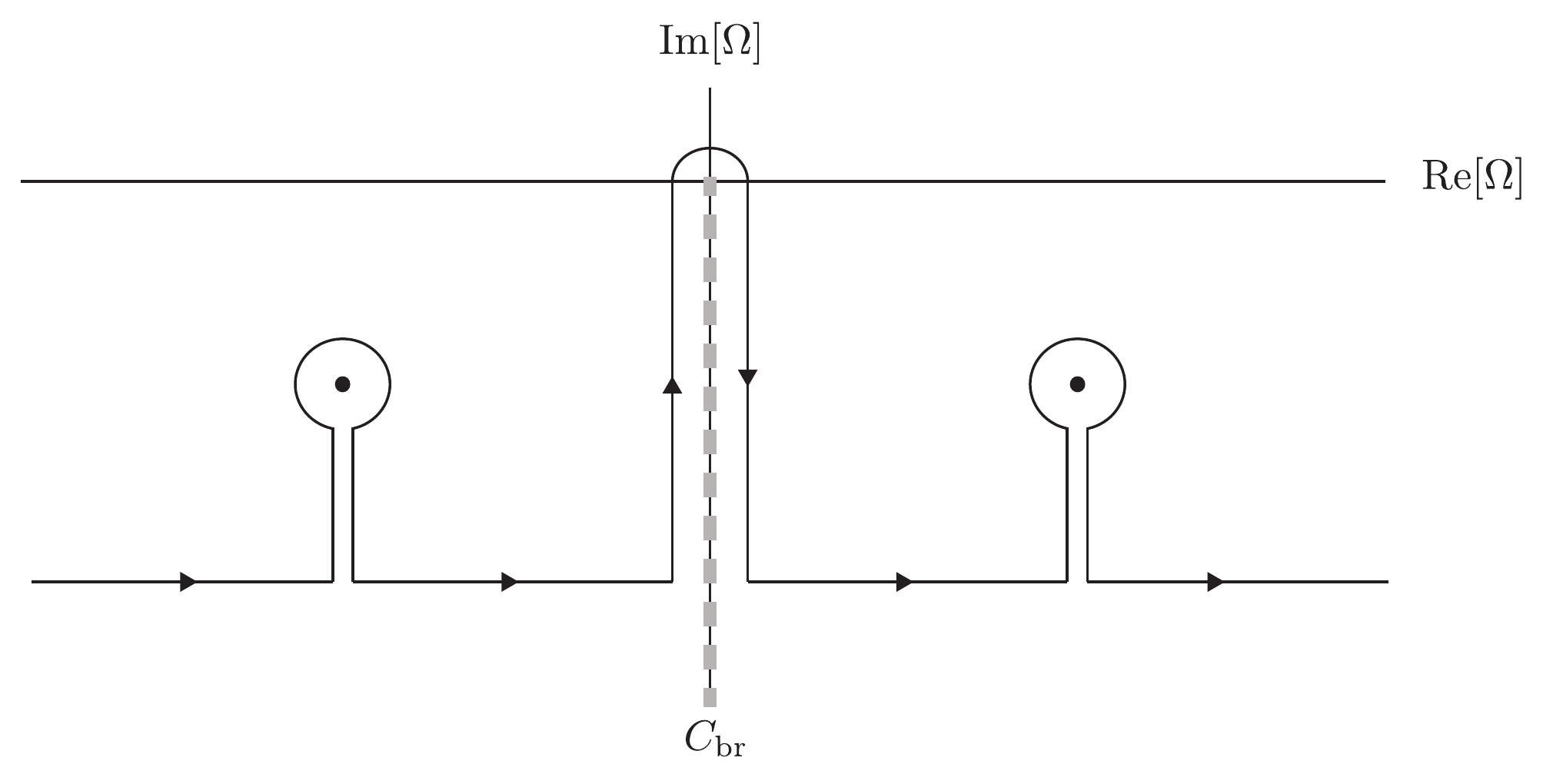}
\caption{Landau contour appearing in Refs.~\citep{sugama-99} and
  \citep{helander-mischenko}.  The branch cut is the grey dashed line labeled
  as $C_{\mathrm{br}}$.} 
\label{Landau-contour-fig}%
\end{figure}

Taking the field-line average of Eq.~(\ref{gk_per}), one can perform the velocity integrals appearing there analytically following 
\citet{biglari}. This results in 
\begin{align}
1 + k_\perp^2 \lambda_D^2 = \frac{1}{2}( D_{+} + D_{-}) \ , \;\;
D_{\pm} = \frac{1}{\sqrt{\pi}} \int 
\frac{\Omega \mp \Omega_*^T}{\Omega \mp \xperp^2/2 \mp \xpar^2} 
\; \exp(-x^2) \, \xperp \df\xperp \df\xpar   %
\label{linear-integral-eq-c}
\end{align}
where we write $\omega_d = \hat{\omega}_d (\xpar^2 + \xperp^2/2)$, $\Omega =
\omega/\hat{\omega}_d$, $x = \sqrt{\xpar^2 + \xperp^2}$, 
$\Omega_*^T = \omega_*^T/\hat{\omega}_d = \Omega_* [1 + \eta(x^2 - 3/2)]$, and $\Omega_* = \omega_*/\hat{\omega}_d$.  
Note that the function $D_{+}$ was obtained by \citet{biglari}, and here we generalize their
calculation to obtain $D_{-}$, which arises because of the sign difference
between the ion (positron here) and the (non-adiabatic) electron drifts. 
To compute this function, one can perform the same integrals, but with the complex frequency in the lower half plane.
We find
\begin{equation}
D_{\pm}(\Omega) = Y_{\pm}^2 + \Omega_* \left\{\left[\pm\frac{\eta-1}{\Omega} -
    2\eta \right] Y_{\pm}^2 \pm 2\eta Y_{\pm}\right\} 
\end{equation}
where
\begin{eqnarray}
&&{} Y_{+}(\Omega) = \int_{ \infty}^{\Omega} \frac{dz}{\sqrt{z}} \exp(z - \Omega) =
-\sqrt{\Omega} Z(\sqrt{\Omega}),\\ 
&&{} Y_{-}(\Omega) = \int_{-\infty}^{\Omega} \frac{dz}{\sqrt{z}} \exp(\Omega - z) =
\sqrt{\Omega}\left[ 2 i \sqrt{\pi} \exp(\Omega) - i Z(-i\sqrt{\Omega})\right], 
\end{eqnarray}
and $Z$ is the plasma dispersion function.  When $\Im[\Omega] > 0$,
$\sqrt{\Omega}$ is defined as the principle root (which lies in the upper half
plane).  To treat Landau damping, $Y_{\pm}$ must be analytically continued to
perform the contour integral in the inversion of the Laplace transform.  In
particular, we must choose the branch of $\sqrt{\Omega}$ so that the function
remains analytic.  Very closely related problems were treated by
\citep{sugama-99} and \citep{helander-mischenko}, who chose the branch cut of
the function $\sqrt{\Omega}$ to lie along the negative imaginary axis.  This
approach allows pole contributions to be picked up in the usual fashion, with
an additional contribution coming from integration along the branch cut; see
Fig.~\ref{Landau-contour-fig}.   

\begin{figure*}
    \centering
    \begin{subfigure}[b]{1. \textwidth}
        \centering
	\includegraphics[width=\textwidth]{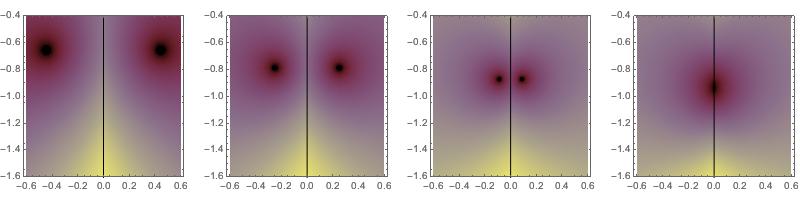}
        \caption{Symmetrized case: branch cut of $\sqrt{\Omega}$ lies on
          negative imaginary axis.  Both poles present until they cross the
          branch cut and are lost.} 
    \end{subfigure}%
     \newline
    \begin{subfigure}[b]{1. \textwidth}
        \centering
	\includegraphics[width=\textwidth]{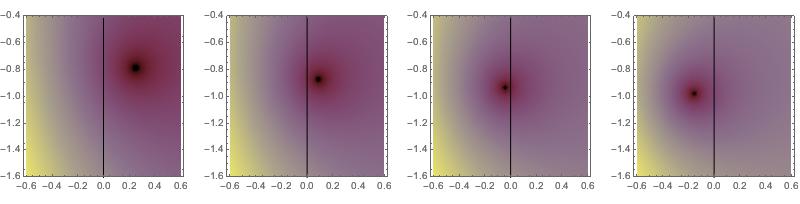}
        \caption{With branch cut of $\sqrt{\Omega}$ placed on negative real
          axis, one pole can be found.} 
    \end{subfigure}%
     \newline
    \begin{subfigure}[b]{1. \textwidth}
        \centering
	\includegraphics[width=1.0\textwidth]{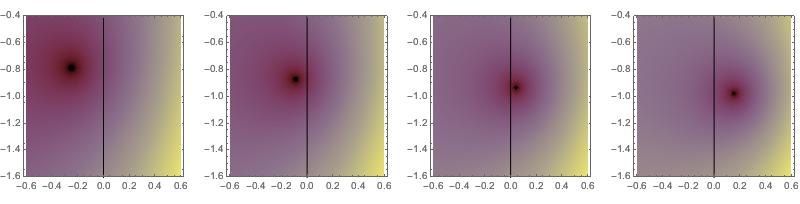}
        \caption{With branch cut of $\sqrt{\Omega}$ placed on positive real
          axis, one pole can be found.} 
    \end{subfigure}%
    \caption{Density plot of dispersion function $1 + \kperp^2\lambda_D^2 -
      (D_{+} + D_{-})/2$ in the complex $\Omega$ plane.  Poles are observed to
      cross the negative imaginary axis as $\kperp^2\lambda_D^2$ is varied.
      Here $\eta = 10$, $\Omega_* = 1$, and the values of $\kperp^2
      \lambda_D^2$ are $6$, $8$, $10$ and $12$.} 
\label{missing-pole-fig}%
\end{figure*}

\begin{figure}
\centering
\includegraphics[width=0.75 \textwidth]{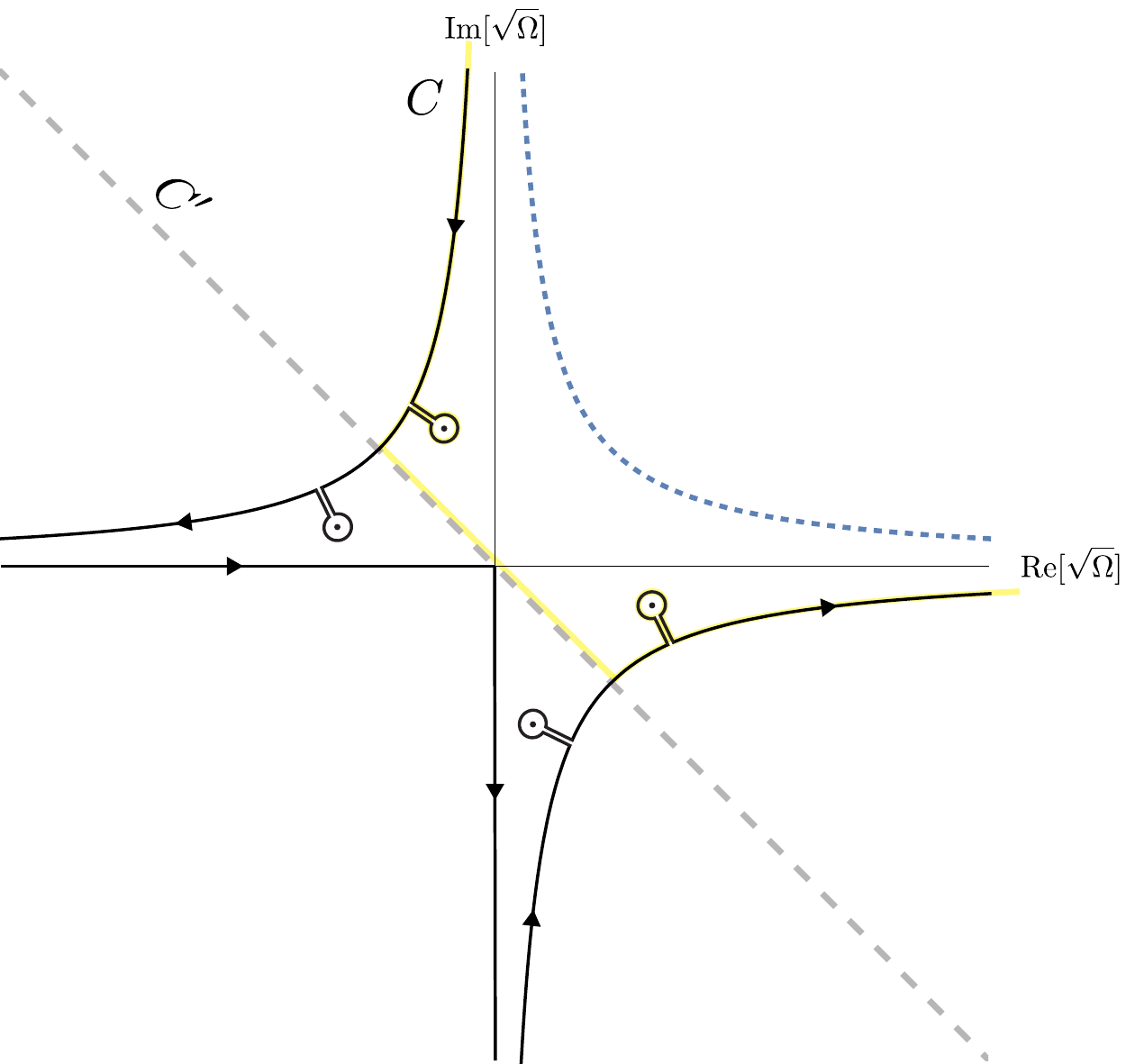}
\caption{Extended Landau contour, $C$.  Note that the negative imaginary
  $\Omega$ axis is mapped to the grey dashed line, labeled $C^{\prime}$, and
  the contour of Fig.~\ref{Landau-contour-fig} is shown in light yellow for
  comparison.  Note that the path of Fig.~\ref{Landau-contour-fig} does not
  encircle the depicted poles that lie beneath $C^{\prime}$, which is why we
  use the modified contour here.} 
\label{Landau-contour2-fig}%
\end{figure}

In the present case we take a different approach, namely to perform the
integral in the $\sqrt{\Omega}$ plane, in which it is possible to continue the
dispersion function analytically over the entire plane, avoiding the need for
a branch cut.  The reason for this choice is that the pole contributions can
cross the negative imaginary axis, and therefore evade detection; a case where
this happens is shown in Fig.~\ref{missing-pole-fig}.  The new contour for
Laplace inversion is shown Fig.~\ref{Landau-contour2-fig}.  In this figure,
the curved paths are contours of constant $\Im[\Omega]$.  Thus, the standard
curve for inverting a Laplace transform is shown in dashed blue.  The new
contour is chosen so that it lies in the upper left and lower right quadrants,
where $\Im[\Omega] < 0$.  The curved portions of the path can be neglected as
compared to the pole contributions, for the usual reason that the
corresponding part of the solution is damped more strongly than the pole
contributions. 

As was found by \citep{sugama-99} and \citep{helander-mischenko}, part of the
integral causes algebraic damping, while the pole contributions, \ie the
complex mode frequencies, can be obtained as the roots of
Eq.~(\ref{linear-integral-eq-c}).  In the present case, the algebraic damping
comes from the integral running along the negative $\Re[\sqrt{\Omega}]$ and
$\Im[\sqrt{\Omega}]$ axes.  The long-time limit of this contribution is
dominated by the $\Omega \rightarrow 0$ point, and goes as $1/t^{2}$.  We
further note that this algebraic contribution correspond to integrals along
the real axis in the $\Omega$ plane.  Thus, the damping can be assured to be
non-exponential, and any exponential damping comes explicitly from the
identified poles. 

In practice, the consequence of the above discussion is that the complex
frequency of Landau-damped modes is determined by roots of the dispersion
function $1 + \kperp^2\lambda_D^2 - (D_{+} + D_{-})/2$ that lie in either the
upper-left and lower-right quadrant in the complex $\sqrt{\Omega}$ plane.
Landau damping will occur at sufficiently large values of $\kperp \lambda_D$.   

At $\kperp \lambda_D \lesssim 1$, the fluid limit $\Omega \gg 1$ can be
applied to Eq.~(\ref{linear-integral-eq-c}), yielding the solution 

\begin{equation}
\omega^2 = \,-\,\frac{\hat{\omega}_d^2}{\kperp^2 \lambda_D^2} \left[ (1 + \eta) \frac{\omega_*}{\hat{\omega}_d} - \frac{7}{4}\right]. \label{z-pinch-fluid-disp-reln}
\end{equation}
Note that the temperature and density gradients act together (via the factor
$1 + \eta$), and a purely density-gradient-driven mode is possible, unlike the
related interchange instability in a conventional electron-ion plasma.  The
mode is predicted to be stabilized when $\hat{\omega}_d$ exceeds $\omega_*$.
This results in the ``fluid'' instability condition (which is the singularity
boundary if $\kperp\lambda_D = 0$ exactly): 

\begin{equation}
\frac{\omega_*}{\hat{\omega}_d}(1+\eta) > \frac{7}{4}.\label{stability-boundary-fluid}
\end{equation}
Note, however, that the threshold condition $\omega = 0$ contradicts the
``fluid'' assumption $\Omega \gg 1$ made above. 
It indicates that higher-order terms may be needed to treat the plasma stability 
at the ``fluid stability boundary'' (singularity boundary for $\kperp\lambda_D = 0$).
Even when $\omega_*$ exceeds $\hat{\omega}_d$ sufficiently for the existence
of an unstable ``fluid'' mode, the mode must succumb to Landau damping for sufficiently
large $\kperp \lambda_D$.  We can take $\Omega \sim 1$ to estimate the
wavenumber where this transition must occur: 

\begin{equation}
\kperp \lambda_D \sim \left| \frac{7}{4} - (1 + \eta) \frac{\omega_*}{\hat{\omega}_d} \right|^{1/2}
\end{equation}
For values of $\kperp \lambda_D$ exceeding this, we return to the Landau damping problem.

\begin{figure*}
    \centering
    \begin{subfigure}[b]{.3 \textwidth}
        \centering
	\includegraphics[width=\textwidth]{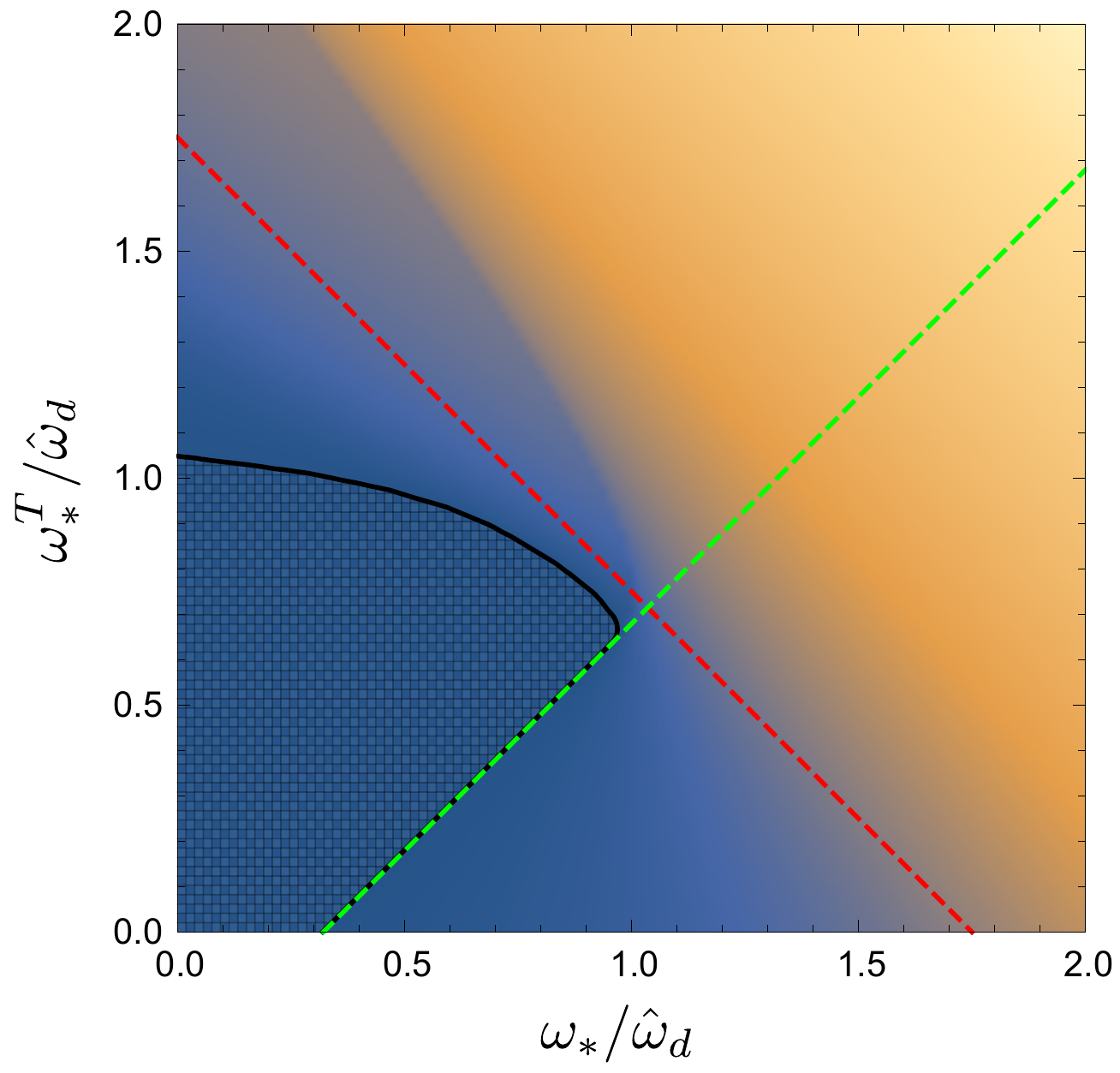}
        \caption{$\kperp^2\lambda_D^2 = 0.01$} 
    \end{subfigure}%
    \begin{subfigure}[b]{.3 \textwidth}
        \centering
	\includegraphics[width=\textwidth]{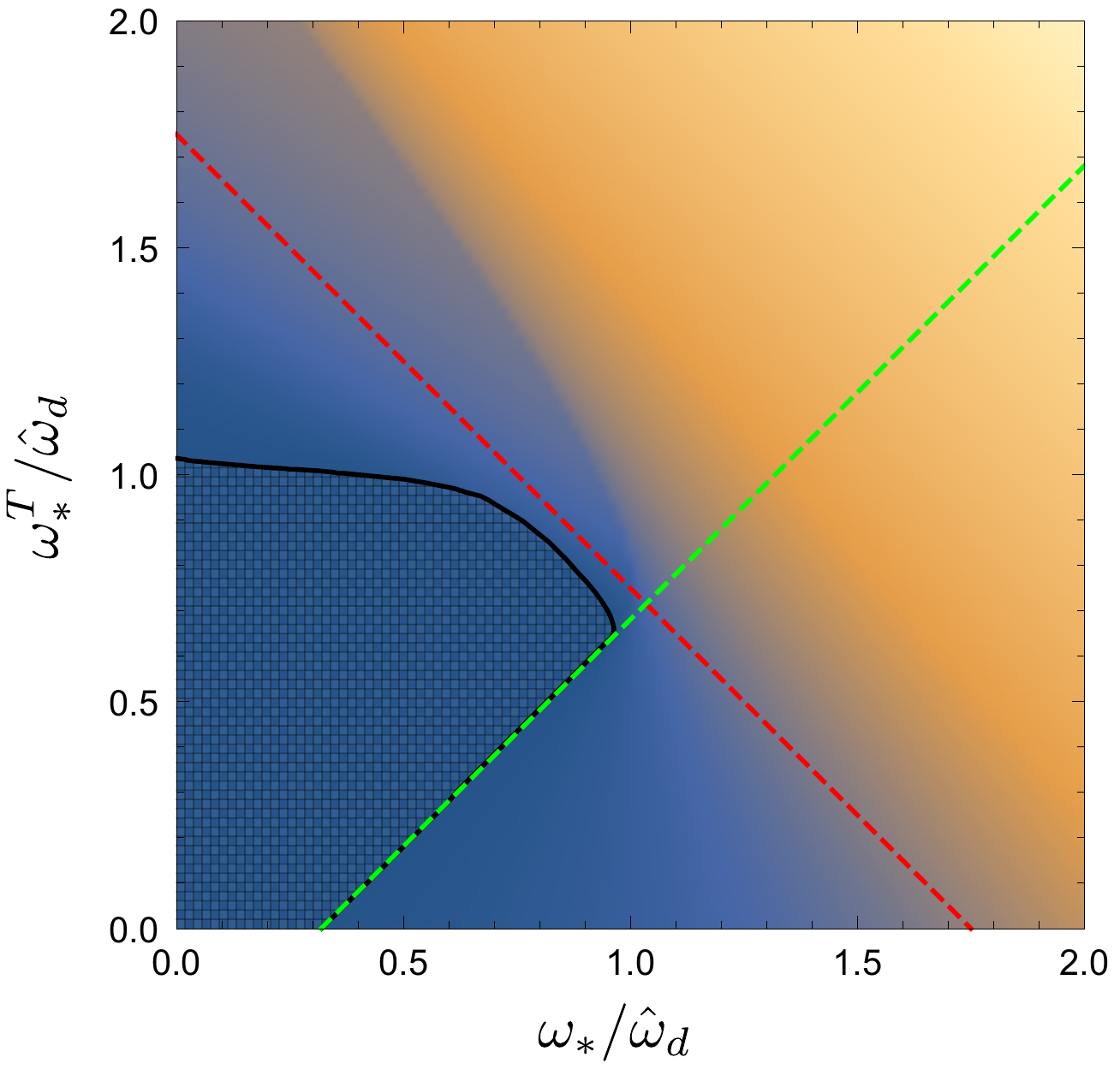}
        \caption{$\kperp^2\lambda_D^2 = 0.0025$} 
    \end{subfigure}%
    \begin{subfigure}[b]{.3 \textwidth}
        \centering
	\includegraphics[width=1.0\textwidth]{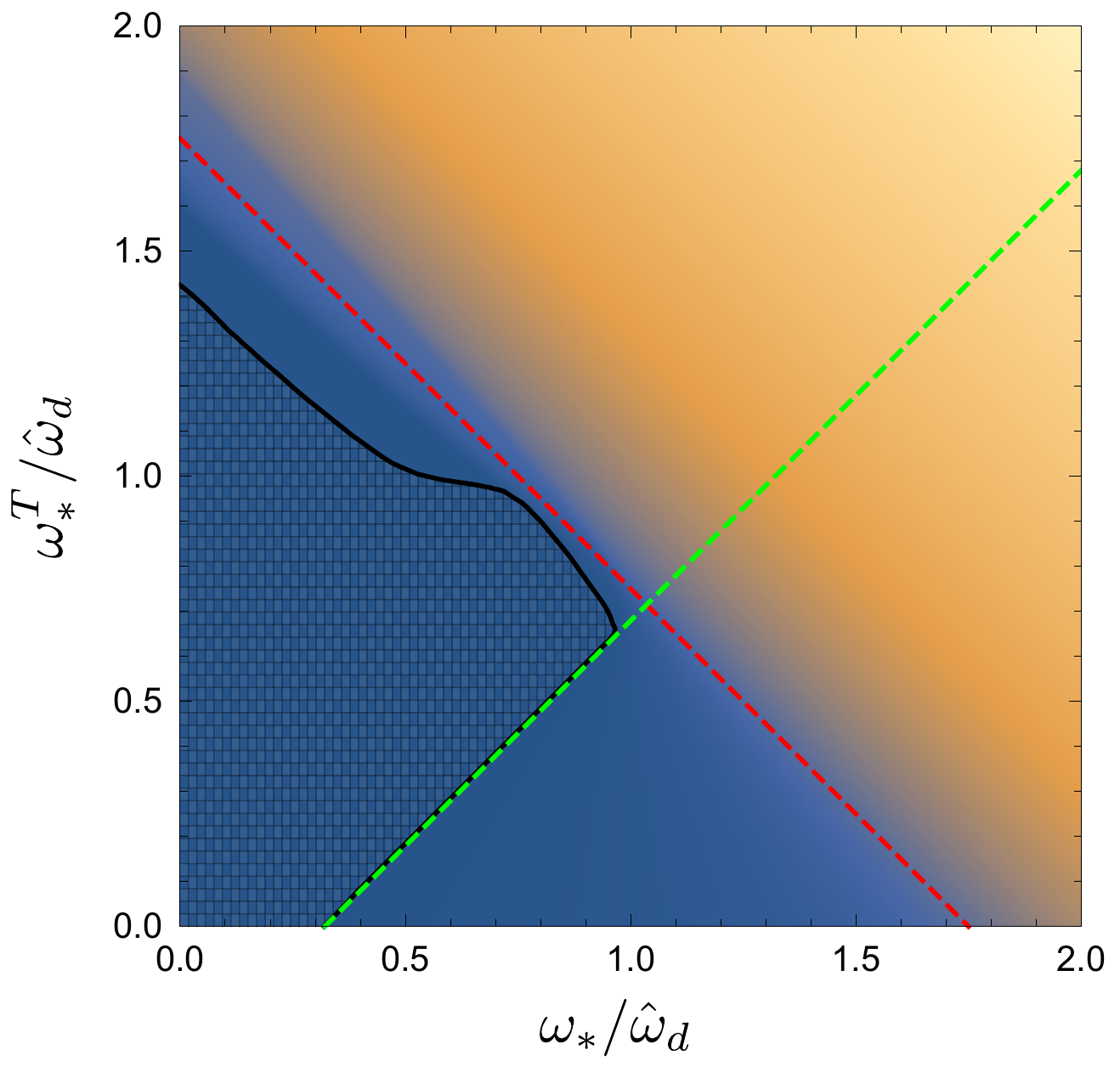}
        \caption{$\kperp^2\lambda_D^2 = 0.000625$} 
    \end{subfigure}%
    \caption{Stability diagram for several values of $\kperp^2\lambda_D^2$, compared with theoretical stability boundaries.
    The colour of the density plot corresponds to
  numerically obtained growth rate (dark blue is zero and large positive values are yellow).  The region of absolute stability is darkened and bordered by a solid black contour.
  The theoretical stability lines (dashed red, and dashed green) correspond respectively to
  Eqs.~(\ref{stability-boundary-fluid}) and (\ref{stability-boundary-res}).  
  Note that the deviation from the red theoretical stability boundary (which does not include finite-$\kperp\lambda_D$ corrections) 
  decreases as $\kperp^2\lambda_D^2 \rightarrow 0$, though surprisingly slowly.} 
\label{stability-diagram-fig}%
\end{figure*}

We can also derive the ``resonant stability boundary'' taking $\Omega
\rightarrow 0$ in Eq.~(\ref{linear-integral-eq-c}), obtaining 

\begin{equation}
\frac{\omega_*}{\hat{\omega}_d}(1- \eta) = \frac{1 +
  \kperp^2\lambda_D^2}{\pi}.\label{stability-boundary-res} 
\end{equation}
As it turns out, the true stability boundary runs along portions each of the
two stability boundaries, Eqs.~(\ref{stability-boundary-fluid}) and
(\ref{stability-boundary-res}), as demonstrated in
Fig.~\ref{stability-diagram-fig}. 


%
%
\section{Point dipole case}\label{sec:dipole}
Now we turn to dipole geometry. In contrast to the Z-pinch limit, the magnetic
field strength and perpendicular wave vector change along field lines in the
dipole geometry, i.~e.~$B = B(\psi,l)$ and $k_{\perp} = k_{\perp}(\psi,l)$,
{\em etc.}, depend on $\psi$ the poloidal flux and $l$ the distance measured
along a field line. 
It has been shown by \citet{kessner} that the bounce-averaged drift frequency in the point
dipole can be approximated with very good accuracy as 
$\ov{\omega}_d  \approx 4 k_{\varphi} m v^2 / (3 e \psi)$.  
Using this approximation, the integrals 
in Eq.~(\ref{gk_per}) can be factorised into velocity ($v$) and
pitch-angle ($\lambda$) parts: 
\be
\label{dipole_factorized}
\Big( 1 + k_{\perp}^2 \lambda_D^2 \Big) \phi = \Lambda \intl_0^{1/B} 
\frac{B \df \lambda}{\sqrt{1 - \lambda B}} \; \bar{\phi} \ , \;\;
\Lambda = \frac{1}{n_0} \intl_0^{\infty} 
\frac{\omega^2 - \ov{\omega}_d \omega_*^T}{\omega^2 - \ov{\omega}_d^2} \, f_0
\, 2 \pi v^2 \df v
\ee
In this Section, we will focus 
on the growing solutions with $\gamma = -i \omega > 0$. 
To begin, we assume $k_{\perp} \lambda_D \rightarrow 0$ and employ the relations: 
\be
\oint \df l \intl_0^{1/B} \frac{\df \lambda}{\sqrt{1 - \lambda B}} \,
\ov{\phi} = 2 \, \oint \frac{\df l}{B} \, \phi \ , \;\; 
\ov{\phi}(\lambda) = \frac{1}{\oint \frac{\df l}{\sqrt{1 - \lambda B}}} 
\oint \frac{\phi \, \df l}{\sqrt{1 - \lambda B}} 
\ee
In this case, the dispersion relation is simply $\Lambda = 1/2$. Expressing
the velocity integrals through the plasma dispersion function $Z_0(\zeta)$, we obtain
\ba
&&{} \Lambda = \frac{1}{2} \left[ \frac{\omega_* \eta}{\omega \zeta} \ov{Z}_6(\zeta) +
  \frac{\omega_*}{\omega \zeta} \left(1 - \frac{3 \eta}{2}\right) \ov{Z}_4(\zeta) -
  \zeta \ov{Z}_2(\zeta) \right] \ , \;\; 
\ohatTwo = \frac{4 k_{\varphi} T}{3 e \psi} 
\\
&&{} \zeta = \sqrt{\frac{\omega}{2 \ohatTwo}} \ , \;\;
\ov{Z}_{n}(\zeta) = Z_{n}(\zeta) + i Z_{n}(i \zeta) \ , \;\;
Z_n(\zeta) = \frac{1}{\sqrt{\pi}} \intl_{-\infty}^{\infty} \frac{x^n e^{-x^2}
  \df x}{x - \zeta}
\ea
which leads to the dispersion relation 
\be
\label{disp_rel_dipole}
\frac{\omega_* \eta}{\omega \zeta} \ov{Z}_6 +
  \frac{\omega_*}{\omega \zeta} \left(1 - \frac{3 \eta}{2}\right) \ov{Z}_4 -
  \zeta \ov{Z}_2 = 1 
\ee
%
\begin{figure}
  \centerline{\includegraphics{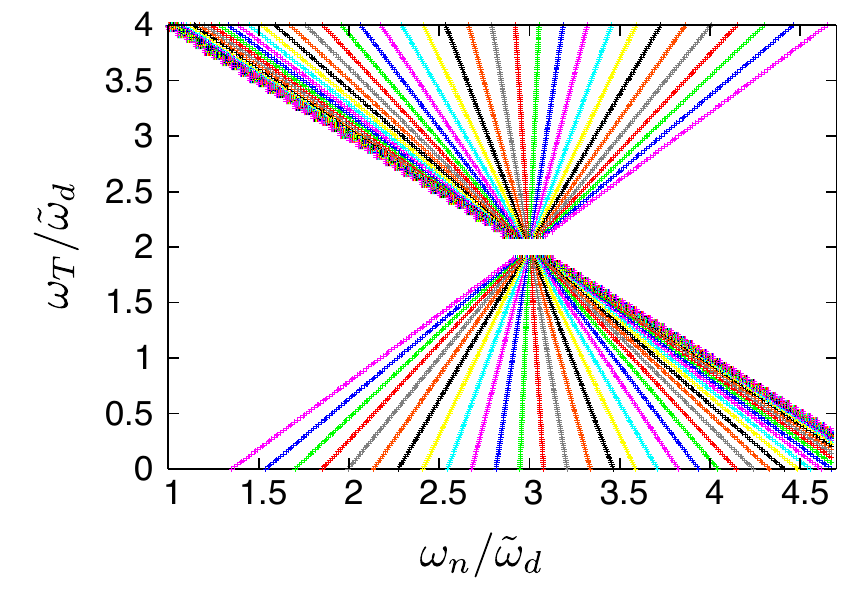}}
  \caption{Contours of constant growth rate $\gamma$ plotted in the two-dimensional parameter
    space $(\osn, \osT)$. The diamagnetic frequencies $\osn$ and
    $\osT$ are normalised to the drift frequency
    $\ohatTwo$. Different colours indicate different values of
    $\gamma$. The density of the contours shows how fast the growth rate
    changes. Note that this density is especially high near the
    ``singularity boundary'', defined in Eq.~(\ref{singularity}).}
\label{fig:dipole}
\end{figure}
Taking 
the limit $\zeta \rightarrow 0$, we find the ``stability'' boundary \citep{per_prl}:
\be
\label{stability}
 \frac{\osn - \osT}{\ohatTwo} = 1.  \ , \;\; \osn = \omega_* \ , \;\; \osT = \eta \omega_*
\ee
Taking the opposite limit $\zeta \rightarrow \infty$,  
we find the ``singularity'' boundary:
\be
\label{singularity}
\frac{\osn + \osT}{\ohatTwo} = 5  \;\; \Leftrightarrow \;\;
\totd{\ln (n T)}{\ln \psi} = \frac{20}{3} 
\ee
Interestingly, the singularity boundary coincides with the MHD stability threshold 
\citep{per_jpp}, although Eq.~(\ref{singularity}) has been obtained within the
electrostatic formalism. Both the stability and the singularity boundaries
can be seen in the numerical solution of the dispersion relation
(\ref{disp_rel_dipole}), shown in Fig.~\ref{fig:dipole}. Here, contours of
constant growth rate $\gamma$ are plotted in the two-dimensional parameter
space $(\osn, \osT)$. 
Different colours in Fig.~\ref{fig:dipole} indicate different values of the growth rate. The
density of the contours shows how fast the growth rate changes. One sees that
the contour density is especially high near the singularity boundary.
%
The numerical solutions of Eq.~(\ref{disp_rel_dipole}) in the domain
bounded by Eqs.~(\ref{stability}) and (\ref{singularity}) represent all unstable modes 
with finite growth rates.

In Fig.~\ref{fig:dipole}, one sees a special point in the parameter space where the stability
line $\osn - \osT = \ohatTwo$ crosses the singularity line
$\osT + \osn = 5 \, \ohatTwo$. The solution of this system of
equations is $\osT = 2 \, \ohatTwo$ and $\osn = 3 \, \ohatTwo$. At this point $\eta = 2/3$ and $\osT/\omega =
1/\zeta^2$. The dispersion relation at the crossing point reduces to the expression:
\be
\label{cross_id}
\frac{\ov{Z}_6}{\zeta^3} - \zeta \ov{Z}_2 = 1 
\ee
We notice, however, that this expression is an identity, i.~e.~ it is valid
for all values of $\zeta$, and can be derived from the definition of
$\ov{Z}_n$.  Thus it cannot be used to determine $\zeta$.  Nevertheless, one
can simplify the dispersion relation by transforming the parameter-space
coordinates $(\osn, \osT)$ so that the origin coincides with the crossing point: 
\be
\label{crossing_point}
\frac{\osT}{\ohatTwo} = 2 + \tau \ , \;\;
\frac{\osn}{\ohatTwo} = 3 + \nu \ , \;\;
\eta = \frac{2 + \tau}{3 + \nu}
\ee
Written in these $(\nu, \tau)$ parameters, the dispersion relation,
Eq.~(\ref{disp_rel_dipole}) reduces to 
\be
\tau \ov{Z}_6 + \left( \nu - \frac{3 \tau}{2} \right) \ov{Z}_4 = 0
\ee
One sees that the contours of constant growth rate (i.~e.~constant $\zeta$ and
therefore constant $\ov{Z}_6/\ov{Z}_4$) are indeed straight lines, in agreement with Fig.~\ref{fig:dipole}:
\be
\frac{\nu}{\tau} = \frac{3}{2} - \frac{\ov{Z}_6}{\ov{Z}_4} \;\;
\Leftrightarrow \;\;
\tau = \left[\frac{3}{2} - \frac{\ov{Z}_6(\zeta)}{\ov{Z}_4(\zeta)}\right]^{-1} \; \nu
\ee

\begin{figure}
  \centerline{\includegraphics[width=0.75 \textwidth]{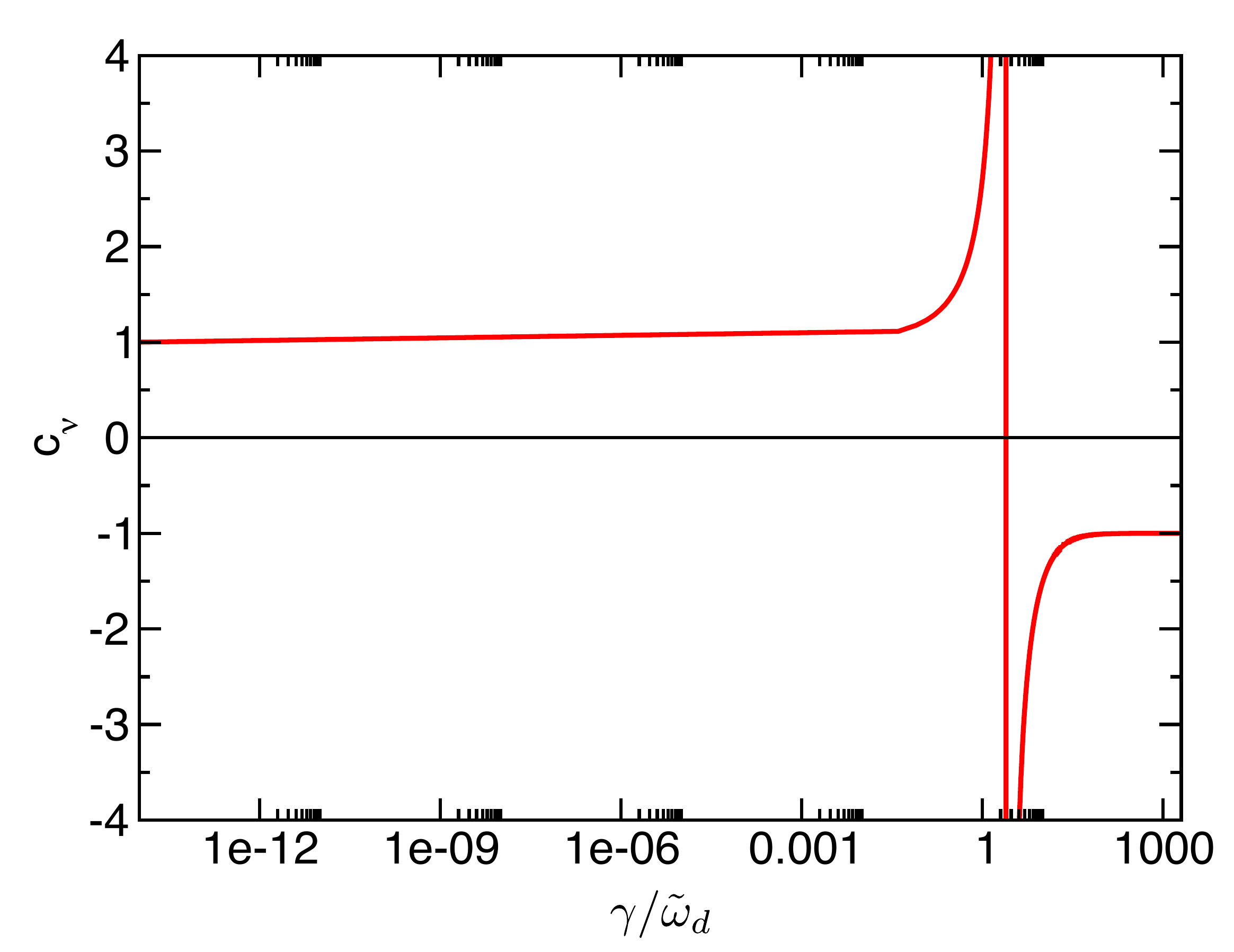}}
  \caption{Frequency-dependent coefficient $c_{\nu}$ as a function of $\zeta$.}
\label{fig:tau_nu}
\end{figure}

Using the identity Eq.~(\ref{cross_id}), one can cast the dispersion relation Eq.~(\ref{disp_rel_dipole}) 
into an alternative form that is particularly simple :
\be
\osT = c_{\nu}(\zeta) \osn + \Big[ 2 - 3 c_{\nu}(\zeta) \Big] \ohatTwo \ , \;\;
c_{\nu}(\zeta) = \left[\frac{3}{2} - \frac{\ov{Z}_6(\zeta)}{\ov{Z}_4(\zeta)}\right]^{-1}
\ee
We stress that this formulation is equivalent to Eq.~(\ref{disp_rel_dipole}). 
Only the algebraic identity Eq.~(\ref{cross_id}) and no
additional assumptions were needed to derive it. The growth rate enters only
through the coefficient $c_{\nu}(\zeta)$ which is plotted in Fig.~\ref{fig:tau_nu}. One
sees that the asymptotic values of this coefficient are
\be
\lim_{\zeta\rightarrow0} c_{\nu}(\zeta) = 1 \ , \;\;
\lim_{\zeta\rightarrow\infty} c_{\nu}(\zeta) = \,-\,1
\ee
It is straightforward to recover the stability and singularity boundaries 
from these values. The crossing point corresponds to the singularity of
$c_{\nu}(\zeta)$, as shown in Fig.~\ref{fig:tau_nu}. 

\subsection{Finite Debye length}
From the preceding results it can be inferred that $k_\perp \lambda_D$ is a
singular limit.  Indeed Eq.~(\ref{z-pinch-fluid-disp-reln}) demonstrates the
existence of a fluid mode whose growth rate varies inversely with $k_\perp
\lambda_D$.  Therefore, we will re-examine the point dipole limit, assuming a
finite Debye length.  This leads to a resolution of the singularity boundary
that was encountered in the case where $k_\perp \lambda_D$ is taken to be
exactly zero.  We return to Eq.~(\ref{dipole_factorized}): 
\be
\label{eq_phi}
\Big( 1 + k_{\perp}^2 \lambda_D^2 \Big) \phi = \Lambda \intl_0^{1/B} 
\frac{B \df \lambda}{\sqrt{1 - \lambda B}} \; \bar{\phi} \ , \;\;
\Lambda = \frac{1}{n_0} \intl_0^{\infty} 
\frac{\omega^2 - \ov{\omega}_d \omega_*^T}{\omega^2 - \ov{\omega}_d^2} \, f_0
\, 2 \pi v^2 \df v
\ee
Multiplying this equation with $\phi$ and integrating it over the field line
$\oint \phi \; (\df l/B) \ldots$, we obtain the energy principle:
\ba
\label{energy_principle}
\oint \frac{\df l}{B} \, \left[ (k_{\perp} \lambda_D)^2 \phi^2 + 
\intl_0^{1/B}\frac{B \df \lambda}{\sqrt{1 - \lambda B}} \Big(\phi -
\bar{\phi}\Big)^2 \right] \,=\, \\
\,=\, \left(\Lambda - \frac{1}{2}\right) \, 
\oint \frac{\df l}{B} \intl_0^{1/B}
\frac{B \df \lambda}{\sqrt{1 - \lambda B}} \; \bar{\phi}^2 \nonumber
\ea
This relation is general and has been derived without further assumptions
other than neglecting the dependence of $\ov{\omega}_d$ on the pitch angle $\lambda$. 
It is convenient to obtain the stability condition in the
fluid limit $\omega \rightarrow \infty$. In this limit
\be
\Lambda - \frac{1}{2} \approx 
\,-\, \frac{3 \, \ohatTwo^2}{2 \omega^2} 
\left[ \frac{\osT + \osn}{\ohatTwo} - 5\right]
\ee
This can be reformulated into the stability condition in the fluid limit:
\ba
\label{fluid_stability:dip}
\omega^2 = \,-\, 
\frac{ 
  3 \, \ohatTwo^2 \oint \frac{\df l}{B} \intl_0^{1/B} \frac{B \df \lambda}{\sqrt{1 - \lambda B}}
  \; \bar{\phi}^2 
}
{
  2 \oint \frac{\df l}{B} \, \left[ (k_{\perp} \lambda_D)^2 \phi^2 +  
\intl_0^{1/B}\frac{B \df \lambda}{\sqrt{1 - \lambda B}} \Big(\phi -
\bar{\phi}\Big)^2 \right] 
}
 \; \left( \frac{\osT + \osn}{\ohatTwo} - 5\right)
\ea
One sees that the plasma is unstable above the singularity line $(\osT + \osn)/\ohatTwo > 5$ 
and the singularity in the growth rate, previously observed for $k_\perp\lambda_D = 0$, is removed. Note that a similar fluid-type instability was found in the Z pinch, see Eq.~(\ref{stability-boundary-fluid}). 

Interestingly, the energy principle can also be used to obtain another useful
result. It follows from the energy principle that $\phi = \ov{\phi}$ if 
$k_{\perp} \lambda_D = 0$, implying $\gav{\phi} = \ov{\phi} = \phi$, i.~e.~$\partial \phi/\partial l = 0$. 
It is a consequence of the energy principle Eq.~(\ref{energy_principle}) and
the dispersion relation $\Lambda = 1/2$, which we have previously shown to
hold for $k_{\perp} \lambda_D = 0$, see Eq.~(\ref{dipole_factorized}). 
If $k_{\perp} \lambda_D$ is finite, but small, the deviation of $\phi$ from
its bounce average or field-line average should also be small $\phi \approx
\gav{\phi} \approx \ov{\phi}$. Introducing the splitting:
\be
\phi = \gav{\phi} + \wt{\phi} \ , \;\; \Bgav{\wt{\phi}} = 0 \ ,\;\; 
\wt{\phi} \sim {\cal O}\left(k_{\perp}^2 \lambda_D^2\right)
\ee
with $\gav{\phi}$ denoting the field-line average and 
$\wt{\phi}$ being small for $k_{\perp} \lambda_D \ll 1$, we can write
Eq.~(\ref{eq_phi}) in the form:
\be
\Big( 1 + k_{\perp}^2 \lambda_D^2 \Big) \gav{\phi} +  \wt{\phi} + k_{\perp}^2
\lambda_D^2 \wt{\phi} = 2 \Lambda \gav{\phi} + \Lambda \intl_0^{1/B} 
\frac{B \df \lambda}{\sqrt{1 - \lambda B}} \; \ov{\wt{\phi}}
\ee
Integrating this equation along the closed field line and taking into account
that 
\be
\oint \frac{\df l}{B} \; \wt{\phi} = 0\ , \;\; 
\oint \frac{\df l}{B} \intl_0^{1/B} \frac{B \df \lambda}{\sqrt{1 - \lambda B}}
\; \ov{\wt{\phi}} = 2 \oint \frac{\df l}{B} \;\wt{\phi} = 0 
\ee
by definition, we obtain the dispersion relation, accurate to the second order:
\be
\Lambda = \frac{1 + \gav{k_{\perp}^2 \lambda_D^2}}{2} + 
{\cal O}\left(k_{\perp}^4 \lambda_D^4\right) \ , \;\;
\gav{k_{\perp}^2 \lambda_D^2} = \frac{1}{\oint \df l/B} \; 
\oint \frac{\df l}{B} \; k_{\perp}^2 \lambda_D^2
\ee
In the crossing-point notation Eq.~(\ref{crossing_point}), this dispersion relation becomes 
\be
\frac{\ov{Z}_4}{2 \zeta^3} \; \left(\nu - \frac{\tau}{c_{\nu}}\right) =
\gav{k_{\perp}^2 \lambda_D^2}.
\ee
It is instructive to find the asymptotes of this dispersion
relation. For $\zeta \rightarrow \infty$ 
\be
\nu + \tau = \,-\, \frac{4 \zeta^4}{3} \; \Bgav{k_{\perp}^2 \lambda_D^2} 
\;\; \Longleftrightarrow \;\; 
\omega^2 = \,-\,\frac{3 \, \ohatTwo^2}{\gav{k_{\perp}^2 \lambda_D^2}} \; 
\left[ \frac{\osT + \osn}{\ohatTwo} - 5 \right]
\ee
One sees that plasma is unstable if 
$\osT + \osn > 5\,\ohatTwo$, in agreement with Eq.~(\ref{fluid_stability:dip}). 
At the singularity line itself, $\omega = 0$ which
contradicts the condition $\zeta \gg 1$, assumed above. This contradiction can be
resolved taking higher-order terms in the plasma dispersion function into account. 
For this sake, we expand the plasma dispersion function \citep{Fried_Conte} as
\ba
&&{} Z_0 = i \sqrt{\pi} e^{-\zeta^2} - \frac{1}{\sqrt{\pi}} \sum_{n=0}^{\infty}
\frac{\G(n+1/2)}{\zeta^{2n + 1}} \,=\, \\
&&{}  \,=\; i \sqrt{\pi} e^{-\zeta^2} - \frac{1}{\zeta} - \frac{1}{2 \zeta^3} -
\frac{3}{4 \zeta^5} - \frac{15}{8 \zeta^7} - \frac{105}{16 \zeta^9} -
\frac{945}{32 \zeta^{11}} \nonumber 
\ea
This expansion leads to the dispersion relation:
\be
\Bgav{k_{\perp}^2 \lambda_D^2} = \,-\,\frac{3}{4 \zeta^4} \left[(\tau + \nu) +
\frac{35}{4 \zeta^4} (3 \tau + \nu) \right] 
\ee
At the singularity line $\tau + \nu = 0$. Here, the dispersion relation degenerates to
\be
\Bgav{k_{\perp}^2 \lambda_D^2} = \,-\,\frac{105 \tau}{8 \zeta^8} 
\;\; \Longleftrightarrow \;\;
\omega^4 = \,-\, \tau \, \frac{210 \, \ohatTwo^4}{\gav{k_{\perp}^2 \lambda_D^2}}
\ee
Note that below the crossing point ($\tau < 0$), there is a single unstable root that is purely growing, whereas above the crossing point ($\tau > 0$), there are two unstable solutions with complex frequencies satisfying ${\rm Im}(\omega) = |{\rm Re}(\omega)|$.
In the opposite limit $\zeta \ll 1$, finite Debye length is less important. It only slightly shifts the stability boundary:
\be
\nu - \tau = \Bgav{k_{\perp}^2 \lambda_D^2} \;\; \Longleftrightarrow \;\;
\frac{\osn - \osT}{\ohatTwo} = 1 + \Bgav{k_{\perp}^2 \lambda_D^2}
\ee

Summarizing, taking finite Debye length into account resolves the singularity in the growth rate appearing in Fig.~\ref{fig:dipole}. 
Instead one finds at the singularity boundary that the mode is purely growing ($\Re[\omega] = 0$) below the crossing point 
and has finite frequency and growth rate above the crossing point. The plasma is thus stable only if $\osn - \osT < \ohatTwo$ and $\osn + \osT < 5\,\ohatTwo$ are both satisfied (i.~e.~ stability is only observed in the left triangular region of Fig.~\ref{fig:dipole}). 
This clarifies the role of the two stability lines identified by \citet{per_prl} and \citet{per_jpp}.  Note that the point-dipole (Fig.~\ref{fig:dipole}) and Z-pinch (Fig.~\ref{stability-diagram-fig}) stability diagrams are similar, with the main difference being simply the location of the crossing point of the stability lines.
%
%
\section{Conclusions}\label{sec:conclusions}

In this paper, we have studied the drift-kinetic stability of a pair plasma, of
equal positron and electron temperature and density, confined by a
dipole magnetic field. The Z-pinch and point-dipole limits have both been
considered, and the resulting dispersion relations have been derived, solved,
and compared. We have found electrostatic instabilities in pure pair plasmas
driven by the magnetic curvature, temperature and density gradients. In
point-dipole geometry, when the Debye length is taken to be exactly zero, we
have found that instabilities exist for the parameters in the domain bounded 
by Eqs.~(\ref{stability}) and (\ref{singularity}). Their growth rate decreases
towards the stability boundary defined by Eq.~(\ref{stability}), and increases
towards the singularity boundary defined by Eq.~(\ref{singularity}).
Visually, this is seen in Fig.~\ref{fig:dipole} as an
increase in the density of the contours of constant growth rate. The
singularity can be resolved taking a small but finite Debye length into
account. This is associated with a fluid-type mode that is absent if
$k_\perp\lambda_D = 0$ exactly. In the Z-pinch limit, the stability diagram
found is similar.  With these 
observations, Figs.~\ref{fig:dipole} and Fig.~\ref{stability-diagram-fig}
describe rather thoroughly the drift-kinetic stability of pair plasmas in
dipole geometry.  We thereby clarify the role of the stability lines in
parameter space, and conclude that both density and temperature gradients
drive instability.  In Z-pinch geometry, we have also treated the Landau
damping problem with a novel integration contour, and found exponential and
algebraic damping solutions related to the drift particle motion. 
The existence of such unstable modes is 
 a collective effect that can provide the background turbulence needed for
an inward particle pinch \citep{Isichenko}. Such a pinch could be very helpful
for pair-plasma creation. 
We plan to address this with a gyrokinetic code in future. 
%

\qquad \\
\qquad \\
{\bf Acknowledgments}
%
We acknowledge Thomas Sunn Pedersen and PAX/APEX experiment team for their
interest to our work.
%
%

\bibliographystyle{jpp}
\bibliography{elpos}

\end{document}